\documentstyle[epsbox,preprint,aps]{revtex}

\begin{document}
\title{Extended Quantum Color Coding}
\author{A.~Hayashi, T.~Hashimoto, and M.~Horibe}
\address{Department of Applied Physics\\
           Fukui University, Fukui 910-8507, Japan}
\draft
\maketitle

\begin{abstract}
The quantum color coding scheme proposed by 
Korff and Kempe \cite{Korff04}
is easily extended so that the color coding quantum system is allowed 
to be entangled with an extra auxiliary quantum system. 
It is shown that in the extended scheme we need only 
$\sim 2\sqrt{N}$ quantum colors to order $N$ objects in large $N$ limit,
whereas $\sim N/e$ quantum colors are required in the original 
non-extended version. The maximum success probability has asymptotics
expressed by the Tracy-Widom distribution of the largest eigenvalue of a 
random GUE matrix.
\end{abstract}


\newcommand{\ket}[1]{|\,#1\,\rangle}
\newcommand{\bra}[1]{\langle\,#1\,|}
\newcommand{\braket}[2]{\langle\,#1\,|\,#2\,\rangle}
\newcommand{\bold}[1]{\mbox{\boldmath $#1$}}
\newcommand{\sbold}[1]{\mbox{\boldmath ${\scriptstyle #1}$}}
\newcommand{\tr}[1]{{\rm tr}\left[#1\right]}
\newcommand{\CH}{{\cal H}}
\newcommand{\CY}{{\cal Y}}

\section{Introduction}

Quantum informational protocols often perform more efficiently than
their classical counterparts when entanglement property of a quantum
system is appropriately utilized. Quantum teleportation \cite{Bennett93}
and quantum dense coding \cite{Bennett92} are typical examples among others. 
Recently Korff and Kempe proposed quantum color coding scheme and 
showed that it is much better than its classical counterpart due to 
entanglement involved in the  color coding  quantum system \cite{Korff04}.

First we describe their classical and then quantum color coding scheme,
slightly modifying
roles of Alice and Bob so that we can naturally extend the scheme later.

Suppose Alice has $N$ boxes containing $N$ distinct classical objects
labeled by an integer from 1 to $N$. In the classical scheme 
Alice paints each box with one of $d$ colors to help her tell,
without opening the boxes,
which box contains which object. Then she passes the boxes
to Bob. Bob is so sloppy that he randomly permutes the boxes and returns
them to Alice. Now Alice's task is to tell which box contains which 
object just by observing the colors of the boxes. Namely Alice should 
guess what permutation was performed by Bob.
Alice's success probability is maximized and given by 
$1/(\frac{N}{d}!)^d$, if she paints equal number of boxes with each 
color (with slight variations if $d$ does not divide $N$).

In the quantum color coding scheme proposed by Korff and Kempe \cite{Korff04}, 
to each box Alice attaches a quantum system on a $d$ dimensional Hilbert space 
$\CH_d$. Before passing the boxes to Bob, 
she prepares a certain color coding state $\ket{\Phi}$ in 
$\CH_d^{\otimes N} \equiv \CH_Q$, which may be entangled among its
subsystems.
The sloppy Bob randomly permutes the boxes and in effect he performs
a unitary permutation operation $T_\sigma$ on $\ket{\Phi}$, 
where $\sigma$ is an element of $S(N)$, the symmetric group of order $N$. 
Receiving the boxes, Alice is to guess the permutation $\sigma$ by
performing some measurements on the permuted state $T_\sigma\ket{\Phi}$.
Alice's strategy therefore consists of the coding state $\ket{\Phi}$ and
POVM on $\CH_Q$. It was shown that the maximum success
probability $p_{\max}(N,d)$ of Alice is given by
\begin{eqnarray}
  p_{\max}(N,d) = \frac{1}{N!}\sum_\rho \min(D_\rho,m_\rho)D_\rho, \label{pmax}
\end{eqnarray}
where $\rho$ indexes irreducible representations of symmetric group 
$S(N)$ with 
dimension $D_\rho$ and $m_\rho$ stands for multiplicity of irreducible
representation $\rho$ on $\CH_d^{\otimes N}$. 
Analyzing asymptotic behavior of representation of $S(N)$ for large $N$,
Korff and Kempe proved the following: \\

\noindent
{\bf Korff-Kempe theorem}\\ \indent
{\it
 Let $r$ be a constant, $d=[rN]$. \\ \indent
1) If $r>1/e$ then $\lim_{N \rightarrow \infty} p_{\max}(N,d)=1$.\\ \indent
2) If $r<1/e$ then $p_{\max}(N,d) \sim \frac{d^N}{N!} \rightarrow 0$ as 
                                 $N \rightarrow \infty$ 
}\\

This theorem implies that we need only $d\sim\!\frac{N}{e}$ quantum 
colors to order $N$ objects, a distinct improvement over the classical
scheme.

In this paper we extend the quantum color coding scheme. In the extended
version Alice has an auxiliary quantum system $R$ on space $\CH_R$
in addition to 
the quantum system $Q$ on $\CH_Q$ attached to the boxes.
Alice prepares a certain state 
$\ket{\Phi^{QR}} \in \CH_Q \otimes \CH_R$, which may be entangled 
in the whole space. Then Alice gives Bob the boxes and system $Q$ attached to
them, but keeps system $R$ at hand. After Bob permutes the boxes and
returns them to Alice, she has a state $T_\sigma^Q \ket{\Phi^{QR}}$,
where $T_\sigma^Q$ is a unitary permutation operator on system $Q$. 
Now Alice's task is to guess the permutation $\sigma$ by performing 
some measurement on the whole system $QR$.
 
In the next section we will show that the maximum success
probability in this extended version is given by
\begin{eqnarray}
  p_{\max}(N,d) = \frac{1}{N!}
          \sum_{\rho(m_\rho \ge 0)} D_\rho^2,
                \label{pmax1}
\end{eqnarray}
where in the summation $\rho$ runs over all inequivalent irreducible
representations appearing on $\CH_Q$.  For small $N$ this may be a small
improvement over the success probability in the non-extended version 
Eq.(\ref{pmax}). According to asymptotic analysis,
however, we can show the following remarkable result in large N limit:\\

\noindent
{\bf Theorem A}\\ \indent
{\it
 Let $r$ be a constant, $d=[r\sqrt{N}]$. \\ \indent
1) If $r>2$ then $\lim_{N \rightarrow \infty} p_{\max}(N,d)=1$.\\ \indent
2) If $r<2$ then $\lim_{N \rightarrow \infty} p_{\max}(N,d)=0$
}\\

\noindent
Thus we need only $d \sim 2\sqrt{N}$ quantum colors to order $N$
objects in the extended quantum color coding scheme.
The detailed asymptotics of $p_{\max}(N,d)$ in the case of $d \sim 2\sqrt{N}$ 
will be discussed later.

\section{Maximum success probability in the extended quantum color coding}
In the extended version of quantum color coding scheme formulated in 
the preceding section, Alice's success probability is written as
\begin{eqnarray}
  p(N,d) = \frac{1}{N!}\sum_{\sigma \in S(N)}
      \bra{\Phi^{QR}} T_\sigma^{Q+} E_\sigma^{QR} T_\sigma^Q \ket{\Phi^{QR}},
                                       \label{pdef}
\end{eqnarray}
where $E_\sigma^{QR}$'s are Alice's POVM elements acting on
$\CH_Q \otimes \CH_R$ and satisfy the POVM conditions:
\begin{eqnarray}
   E_\sigma^{QR} \ge 0,\ \ \ \sum_{\sigma \in S(N)} E_\sigma^{QR} = 1.
                                       \label{Econditions}
\end{eqnarray}

For a given set of $N$ and $d$ we maximize the success probability $p(N,d)$
with respect to the code state 
$\ket{\Phi^{QR}}$ and the set of POVM elements $E_\sigma^{QR}$.
In spite of additional degrees of freedom due to the auxiliary system $R$
introduced in the extended version, 
the argument proceeds quite similarly to the one given in \cite{Korff04}.

First we define an operator 
\begin{eqnarray}
 E^{QR} \equiv \frac{1}{N!}\sum_{\sigma \in S(N)}
               T_\sigma^{Q+} E_\sigma^{QR} T_\sigma^Q,
\end{eqnarray}
so that the success probability $p(N,d)$ defined by Eq.(\ref{pdef})
can be written as
\begin{eqnarray}
   p(N,d) = \bra{\Phi^{QR}} E^{QR} \ket{\Phi^{QR}}.  \label{pdef1}
\end{eqnarray}
This operator $E^{QR}$ satisfies the following conditions:
\begin{eqnarray}
  E^{QR} \ge 0,\ \ \ 
  \sum_{\sigma \in S(N)} T_\sigma^{Q} E^{QR} T_\sigma^{Q+}=1.
                             \label{Econditions1}
\end{eqnarray}
In this equation, positivity of $E^{QR}$ is evident
and the second relation can be shown by the use of the property 
$T_\sigma^Q T_\tau^Q = T_{\sigma\tau}^Q$ and the completeness of
$E_\sigma^{QR}$.

Conversely if an operator $E^{QR}$ satisfies conditions
Eq.(\ref{Econditions1}), then operators defined as
$E_\sigma^{QR} \equiv T_\sigma^{Q} E^{QR} T_\sigma^{Q+}$ satisfy
the POVM conditions Eq.(\ref{Econditions}) and the success probability 
Eq.(\ref{pdef1}) becomes Eq.(\ref{pdef}) with these $E_\sigma^{QR}$'s.
Our task is therefore to maximize the success probability 
$p(N,d)$ expressed as Eq.(\ref{pdef1}) under the conditions 
Eq.(\ref{Econditions1}) on $E^{QR}$.

Let us write the spectral decomposition of the positive operator $E^{QR}$ as
\begin{eqnarray}
  E^{QR} = \sum_{e} \ket{e}\bra{e},
\end{eqnarray}
where states $\ket{e}$'s are orthogonal to each other 
but not necessarily normalized. Then we immediately see that
\begin{eqnarray}
  p(N,d) = \sum_{e} |\braket{\Phi^{QR}}{e}|^2 
        \le \braket{e_{\max}}{e_{\max}},            \label{pbraket}
\end{eqnarray}
where equality holds when the normalized code state $\ket{\Phi^{QR}}$
is taken to be proportional to $\ket{e_{\max}}$ that is the state with
the largest norm among all $\ket{e}$'s.

In order to study the second condition in Eq.(\ref{Econditions1})
we introduce an orthonormal basis for the space $\CH_Q$:
\begin{eqnarray}
    \ket{\rho,b,a}^Q,
\end{eqnarray}
where $\rho$ indexes inequivalent irreducible representations of $S(N)$,
$a$  indexes states belonging to an irreducible representation, 
and other quantum numbers that distinguish equivalent representations 
are collectively denoted by $b$. With an orthonormal basis
$\{\ket{r}^R\}$ for the auxiliary space $\CH_R$, we write 
\begin{eqnarray}
    \ket{\rho,b,a,r} = \ket{\rho,b,a}^Q \otimes \ket{r}^R.
\end{eqnarray}

In this basis we calculate
matrix elements of the second condition in Eq.(\ref{Econditions1}),
$\sum_{\sigma} T_\sigma^{Q} E^{QR} T_\sigma^{Q+}=1$,
and find
\begin{eqnarray}
  \delta_{bb'}\delta_{rr'}
 &=& \frac{N!}{D_\rho} 
        \sum_a \bra{\rho,b,a,r}E^{QR}\ket{\rho,b',a,r'},
                            \nonumber \\
 &=& \frac{N!}{D_\rho} 
        \sum_{a,e} \braket{\rho,b,a,r}{e}\braket{e}{\rho,b',a,r'},
                            \label{Econdition2}
\end{eqnarray}
where $D_\rho$ is the dimension of irreducible representation $\rho$.
In deriving the above relation we used the orthogonal conditions
of representation matrices
$ D_{a_1a_2}^\rho(\sigma)
  \equiv \bra{\rho,b,a_1,r}T_\sigma \ket{\rho,b,a_2,r}$:
\begin{eqnarray}
  \sum_{\sigma \in S(N)} D_{a_1a_2}^{\rho *}(\sigma)
                         D_{a_1'a_2'}^{\rho'}(\sigma)
 = \delta_{\rho\rho'}\delta_{a_1a_1'}\delta_{a_2a_2'} \frac{N!}{D_\rho}.
\end{eqnarray}

Here for each $\rho$ we introduce a rectangular matrix $u^\rho$ by
\begin{eqnarray}
   u_{a,e\,:\,b,r}^\rho \equiv \sqrt{\frac{N!}{D_\rho}} \braket{e}{\rho,b,a,r},
\end{eqnarray}
where rows are indexed by the set $(a,e)$ and columns by the set $(b,r)$. 
In terms of 
the matrix $u^\rho$ the condition Eq.(\ref{Econdition2}) takes the form:
\begin{eqnarray}
   \delta_{bb'}\delta_{rr'} = \sum_{a,e} u^{\rho*}_{a,e\,:\,b,r}
                                         u^\rho_{a,e\,:\,b',r'},
                                    \label{Econdition3}
\end{eqnarray}
implying that the set of all column vectors in the matrix $u^\rho$
is orthonormal.

On the other hand $\braket{e_{\max}}{e_{\max}}$ in Eq.(\ref{pbraket})
is written as
\begin{eqnarray}
   \braket{e_{\max}}{e_{\max}} =
             \sum_\rho \frac{D_\rho}{N!} \sum_{a,b,r}
              |u^\rho_{a,e_{\max}\,:\,b,r}|^2.    \label{emaxnorm2}
\end{eqnarray}

Since all column vectors of $u^\rho$ are normalized we find
$\sum_{a,b,r} |u^\rho_{a,e_{\max}\,:\,b,r}|^2 \le \sum_{b,r}1=m_\rho|R|$,
where $m_\rho=\sum_b1$ is the multiplicity of irreducible representation 
$\rho$ and we denote the dimension of $\CH_R$ by $|R|=\sum_r1$.
It should be noticed that we can always enlarge the rectangular matrix 
$u^\rho$ to a square unitary matrix by appending some appropriate column 
vectors.  Since all row vectors of the enlarged matrix $u^\rho$ are now
normalized, we have
$\sum_{a,b,r} |u^\rho_{a,e_{\max}\,:\,b,r}|^2 \le \sum_a 1= D_\rho$.
Combining these two inequalities we find
\begin{eqnarray}
   \sum_{a,b,r} |u^\rho_{a,e_{\max}\,:\,b,r}|^2 \le 
                 \min(m_\rho|R|,D_\rho).   \label{sumbound}
\end{eqnarray}

It is clear that this upper bound can be attained by some $u^\rho$ under
the conditions Eq.(\ref{Econdition3}), which is the only constraint on 
$u^\rho$. There are infinitely many possibilities for choice of $u^\rho$
and consequently of $E^{QR}$. A simple one would be that a
$\min(m_\rho|R|,D_\rho)$ by $\min(m_\rho|R|,D_\rho)$ submatrix of
$u^\rho_{a,e_{\max}:b,r}$ is taken to be a unit matrix, which leads to
$E^{QR}$ quite analogous to the one given in \cite{Korff04}.

Through Eq.(\ref{pbraket}), Eq.(\ref{emaxnorm2}) and Eq.(\ref{sumbound}),
we conclude that
the maximum success probability $p_{\max}(N,d)$ is given by
\begin{eqnarray}
  p_{\max}(N,d) = \frac{1}{N!} \sum_\rho \min(m_\rho|R|,D_\rho)D_\rho. 
                  \label{pmax2}
\end{eqnarray}

If we set $|R|=1$, we reproduce Eq.(\ref{pmax}), the result in the case 
of the non-extended quantum color coding. 
If Alice has a sufficiently large auxiliary space
$\CH_R$, we have Eq.(\ref{pmax1}) for the maximum success probability
as promised:
\begin{eqnarray}
  p_{\max}(N,d) = \frac{1}{N!} \sum_{\rho(m_\rho \ge 1)} D_\rho^2
              = \frac{1}{N!} \sum_{\rho \in \CY_N(c_1(\rho) \le d)} D_\rho^2, 
                 \label{pmax11}
\end{eqnarray}
where $\rho$ runs over irreducible
representations with $m_\rho \ge 1$, 
or equivalently $\rho$ runs over all 
Young diagrams $\CY_N$ with the length of its first column $c_1(\rho)$ 
not exceeding $d$. 

For small N the maximum success probability 
given in the above Eq.(\ref{pmax11}) is only a small improvement over 
the one in the non-extended scheme. In fact in the case of $N=3$ and $d=2$
, the two maximum success probabilities give the same value 
since $m_\rho \ge D_\rho$ for representation $\rho$ on $\CH_2^{\otimes 3}$.
For large $N$, however, the asymptotics given in the next section shows a 
substantial improvement.
 
\section{Asymptotic success probability for large $N$}
The maximum success probability $p_{\max}(N,d)$ of Eq.(\ref{pmax11}) is a
partial sum of the Plancherel measure $\mu_N(\rho) \equiv \frac{D_\rho^2}{N!}$.
The total sum $\sum_{\rho \in \CY_N} \mu_N(\rho)$ over all irreducible 
representations
is equal to 1 according to the general theory of group representation
\cite{Hamermesh62}.
Therefore if $N$ is finite and $d<N$, $p_{\max}(N,d)$ is 
strictly less than 1, 
since Young diagrams with the length of its first column greater than $d$ 
do not appear on $\CH_Q = \CH_d^{\otimes N}$. But what is asymptotic 
behavior of $p_{\max}(N,d)$ in large $N$ limit?

Asymptotic representation theory of the symmetric group has been extensively
studied (see \cite{Kerov03} for example). Vershik and Kerov showed that 
if properly scaled Young diagrams equipped with the Plancherel measure 
converge to a universal shape as $N$ goes to infinity 
\cite{Vershik85,Kerov03}. 
In particular $r_1(\rho)$, the length of the first row of 
Young diagram $\rho$, has asymptotics $r_1 \sim 2\sqrt{N}$. 
More precisely their theorems state that for any $\epsilon > 0$,
\begin{eqnarray}
   \lim_{N \rightarrow \infty}
    \mu_N \left\{ \rho \in \CY_N : 
             \left| \frac{r_1(\rho)}{2\sqrt{N}}-1 \right| < \epsilon
          \right\} = 1.
\end{eqnarray}
 
In terms of the Plancherel measure the maximum success probability 
in Eq.(\ref{pmax11}) can be written as
\begin{eqnarray}
     p_{\max}(N,d) = \mu_N \{\rho \in \CY_N : r_1(\rho) \le d \},
\end{eqnarray}
due to the symmetry of the Plancherel measure with respect to 
exchange of rows and columns. Now it is readily shown that if 
$d \ge r\sqrt{N}$ for a constant $r>2$ then 
$\lim_{N \rightarrow \infty} p_{\max}(N,d) = 1$ and if
$d \le r\sqrt{N}$ for a constant $r<2$ then
$\lim_{N \rightarrow \infty} p_{\max}(N,d) = 0$.
Thus we obtain Theorem A given in the introduction.

How does $p_{\max}(N,d)$ behave when $d \sim 2\sqrt{N}$?
To answer this question we need a distribution function of $r_1$ 
in a finer scale. According to the Baik-Deift-Johansson theorem 
\cite{Baik9900}, the distribution of $r_1$ 
, if properly scaled and centered, converges to 
the Tracy-Widom distribution as $N \rightarrow \infty$:
\begin{eqnarray}
&& \mbox{for all}\ x\in \bold{{\rm R}} \nonumber \\
&& \lim_{N \rightarrow \infty} \mu_N \left\{
       \rho \in \CY_N :
       \frac{r_1(\rho)-2\sqrt{N}}{N^{1/6}} \le x \right\} 
    = F_{{\rm TW}}(x).
\end{eqnarray}
The Tracy-Widom distribution first arose as the distribution of the 
largest eigenvalue of a random GUE matrix \cite{Tracy94} and 
its (integrated) distribution function $F_{{\rm TW}}(x)$ satisfies
$F'_{{\rm TW}}(x)>0$, $\lim_{x \rightarrow \infty} F_{{\rm TW}}(x) = 1$ and
$\lim_{x \rightarrow -\infty} F_{{\rm TW}}(x) = 0$.

Using the Baik-Deift-Johansson theorem, we immediately obtain the following 
asymptotics for the maximum success probability in the extended quantum color
coding scheme:\\

\noindent
{\bf Theorem B}
\begin{eqnarray}
  p_{\max}(N,d) \sim F_{{\rm TW}}\left( 
           \frac{d-2\sqrt{N}}{N^{1/6}} \right)\ \ \ \ 
           {\rm as}\ N \rightarrow \infty.
\end{eqnarray}

\section{Concluding remarks}
In the non-extended quantum color coding scheme, the code state 
$\ket{\Psi^Q}$ of system $Q$ can be entangled among its subspace $\CH_d$'s.
That is one of the reasons why this quantum scheme performs much better 
than its classical counterpart \cite{Korff04}. 
In this paper we obtained further improvement on
efficiency of the scheme by introducing more entanglement:
the entanglement between system $Q$ and an auxiliary system $R$.
We have again seen an example where entanglement plays an important role 
in quantum protocols.  

As pointed out by Korff and Kempe, their result Eq.(\ref{pmax})
is completely general and can be formulated for any group $G$.
Namely Alice's maximum success probability of guessing Bob's
operation $T_g$, where $g \in G$, is given by
\begin{eqnarray}
  p_{\max}(N,d) = \frac{1}{|G|}\sum_\rho \min(D_\rho,m_\rho)D_\rho, 
                      \label{pmaxG}
\end{eqnarray}
where $|G|$ is the order of the group $G$.

We note that the same thing holds for the extended version 
considered in this paper. If Alice has a sufficiently large auxiliary 
space, her maximum success probability is given by
\begin{eqnarray}
  p_{\max}(N,d) = \frac{1}{|G|}\sum_{\rho(m_\rho \ge 1)} D_\rho^2.
                      \label{pmaxG1}
\end{eqnarray}

This suggests that the extension considered in this paper 
is useful not only for the symmetric 
group but has many other interesting applications.

\end{document}